\documentclass[twocolumn]{aastex701}
\usepackage{ulem}
\usepackage{float}
\usepackage{placeins}
\usepackage{multirow}
\usepackage{longtable}
\newcommand{\kms}{\mbox{km\,s$^{-1}$}}
\newcommand{\etal}{\mbox{\rm{et al.}~~}}
\newcommand{\HI}{\mbox{H\,{\sc i}}}
\newcommand{\K}{$K_s$}

\begin{document}

\title{Peculiar velocities at low Galactic latitude}

\author{Jeremy Mould,$^{1,4} $ H\'el\`ene M. Courtois,$^2$ Ren\'ee C. Kraan-Korteweg,$^3$ and Amber Hollinger$^2$}
\affiliation{$^1$Centre for Astrophysics \& Supercomputing, Swinburne University, Hawthorn, VIC 3122, Australia,\\
$^2$Universit\'e Claude Bernard Lyon 1, IUF, IP2I Lyon, 4 rue Enrico Fermi, 69622 Villeurbanne, France,\\
$^3$Department of Astronomy, University of Cape Town, Private Bag X3, Rondebosch 7701, South Africa\\
$^4$ARC Centre of Excellence in Dark Matter Particle Physics, Australia 3010}
\email[show]{jmould@swin.edu.au}  



\begin{abstract}
The Laniakea Supercluster is the closest 
large scale structure of galaxies. Is such a structure expected in the standard cold dark matter model of cosmology? This would be a relatively simple question to answer, were it not for the fact that the Zone of Avoidance (ZOA) runs right through it. Recent improvements 
to this paucity of data in the innermost ZOA can be made  from systematic 21 cm surveys using the MeerKAT telescope (e.g. Kraan-Korteweg et al. 2024), and implementing these \HI-redshifts as an extension to the CosmicFlows4 database for reconstruction (Hollinger et al. 2026). In this paper we test the assumption 
that for the purpose of reconstruction, additional \HI\ detected galaxies without peculiar velocity determinations could be placed at their Hubble distances. We present infrared photometry of 163 of these in \HI\ detected MeerKAT ZOA galaxies, in addition to 214 
2MASS Extended Sources in the ZOA to determine their peculiar velocities.
Averaging these peculiar velocities into redshift bins, we find 
that peculiar velocity corrections in
the Laniakea Supercluster ZoA region are not prohibitively large, and that one can
proceed with its reconstruction using the  copious redshift data now available.
\end{abstract}

\keywords{
large scale structure — infrared photometry — galaxy magnitudes— scaling relations}

\section{Introduction}
Cosmography has as its definitive goal to describe where galaxies 
and dark matter are
located in the low redshift universe (Courtois \etal 2013, 2025, Dupuy \& Courtois 2023). But the subject impinges on cosmology
in a number of ways, for example, characterizing the structure on small and large scales, the current power spectrum of the density and velocity fluctuations that have
been evolving since the beginning of the matter dominated era, and the relation
of the distribution of matter relative to the distribution of galaxies  
(Kolatt, Dekel \& Lahav 1995; Shaya \etal 2022). 

Mould (2026) has explored the effect of a mass losing cosmology
associated with primordial black holes as dark matter, in which (1) density peaks
arise before matter radiation-equality, (2) matter-radiation equality
occurs earlier than for $\Lambda$CDM and (3) $\Omega_m$ was higher
for z $\gtrsim$ 10 than it is today. On large scales with
b as the bias factor the power spectrum 

$$P(k) \Rightarrow P(k)\left[1 + 2 \frac{\Omega^{4/7}}{3b} + \frac{\Omega^{8/7}}{5b^2}\right]$$
(Feldman, Kaiser \& Peacock 1994), and the increased early $\Omega ~(=~ 
\Omega_m$) supports larger coherent structures and larger velocities of these structures. Establishing a bulk flow in the nearby Universe of the size found in the recent reconstruction by Hollinger \etal (2026) would be an indicator that $\Omega_m$ may have been time varying. Other solutions to the $\Lambda$CDM bulk flow tension may be forthcoming. In this paper we do no more than examine the tension's dependence on Galactic plane extinction and peculiar velocities.

There
are significant impediments to these large scale structure goals, such as observational biases (Nusser 2025)
in the data and also 
large volumes of the Universe that remain uncharted because of the obscuration by the disk of the Milky Way, the so-called Zone of Avoidance (ZOA).

The campaign to deal with the second of these issues has been reviewed by Kraan-Korteweg (2005) and is ongoing. An additional motivation for the work is the fact that, as chance would have it, there is a massive structure of interconnected rich clusters, called Quipu, that is bisected  by the Galactic plane (B\"ohringer \etal 2025). The earlier identified Vela supercluster located behind the ZoA (VSCL; Kraan-Korteweg et al. 2017)  may well form a continuation of the 400 Mpc Quipu structure. Other more well-known large-scale structures in the southern ZOA that remain poorly mapped  at low Galactic are the Great Attractor/Lanikea (Dressler \etal 1987, Tully \etal 2014), the Local Void (Tully 2008, Kraan-Korteweg et al. 2008, Tully et al. 2019, Kurapati et al. 2024), with the Ophiuchus Supercluster in its background (e.g. Wakamatsu et al. 1997, Hasegawa et al. 2000, Louw et al. 2026), and the Perseus Pisces Supercluster in the northern ZOA (e.g. Focardi 1984, Chamaraux et al. 1990, Ramatsoku et al 2016, 2020).

The field is assisted by advances in technology, such as  multi-parameter model fitting and data visualization combined with recent deep systematic interferometric surveys in neutral hydrogen such as Wallaby with ASKAP (Koribalski et al. 2000), and the SARAO MeerKAT radio telescope. With regard to the ZOA, the extraction of \HI-emission of galaxies in the SARAO MeerKAT Galactic Plane Survey (SMGPS; Goedhart et al. 2024) is of particular interest (Kraan-Korteweg et al. 2026).

In this paper we present infrared photometry of a fraction of the galaxies recently detected at 21 cm in the ZOA that are visible in the infrared \K-band, 2.2$\mu$m ($\S$2). In $\S$3 we use the infrared Tully-Fisher relation (Tully \& Fisher 1977, Aaronson \etal 1979, Masters \etal 2014) 
to measure redshift independent distances. We then discuss the question ($\S\S$4,5)  whether 
the prominence of structures behind the Galactic plane  can be divined by using redshifts as distances, or whether a more sophisticated kinematic model is required to get a correct impression of the hidden large-scale structures. 
Our conclusion, presented in $\S$6 
is that the available data allow us to place an upper limit on the peculiar velocities in such a model.
\begin{figure}
\hspace{-.85cm}
\includegraphics[clip, trim=5.65cm 11cm 0.5cm 11cm, width=0.85\textwidth]  {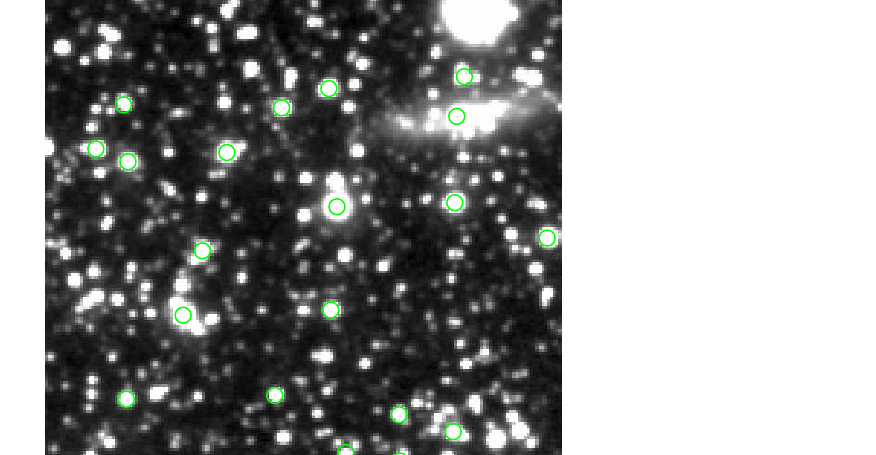}
    \caption{\K-band image of 130211-640338, illustrating the level of stellar crowding in these ZOA fields. 
    The size of the field is {1' $\times$ 1'}, North is up and East to the left. 2MASS stars are circled in green. The galaxy is the 
    horizontally elongated object toward the SE corner with, the 2MASS ID close to its nucleus. }
    \end{figure}

\section{Sample selection}

Our target selection for \K-band photometry of \HI-detected ZOA galaxies is based on recent systematic MeerKAT surveys in the ZOA, 
i.e. a pilot project with MeerKAT16 in Vela (Steyn 2023), \HI-surveys extracted from the SARAO MeerKAT Galactic Plane Survey (SMGPS; Goedhart et al. 2024, Kraan-Korteweg et al. 2026), namely Steyn et al. 2024 (Great Attractor Wall crossing), Kurapati et al. 2024 (Local Void), Rajohnson et al. 2024a (Vela SCL), as well as an extension to slightly higher latitudes (Rajohnson et al. 2024b). We also included Parkes Multi Beam HIZOA detections (Staveley-Smith \etal 2016), its northern extension (Donley \etal 2006) and the Galactic Bulge extension (Kraan-Korteweg \etal 2008), and higher-resolution Parkes observations of lower SNR \HI-profiles (Said \etal 2016a). 
Other SMGps focused on the Crux region (see Steyn \etal 2026), and the Ophiuchus supercluster region just beyond the Local Void (Louw \etal 2026).

\begin{figure}[H]
\vspace{-1.25in}
\hspace{-1in}
\includegraphics[
width=1.8\columnwidth]{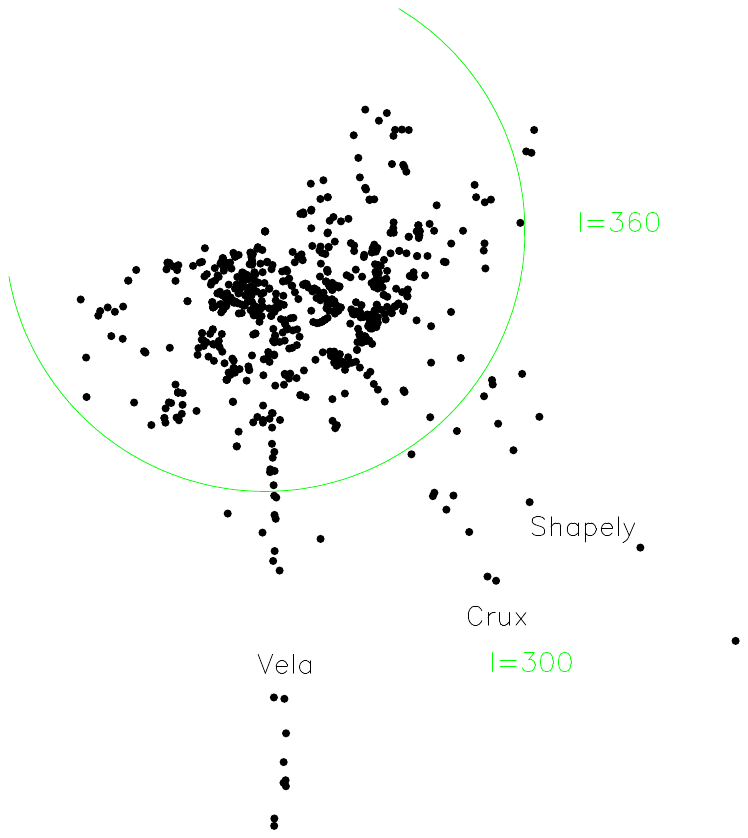}
\vspace{-0.9in}
\caption{
The view of the data from the Galactic pole. The Sun is at the centre and the green circle delineates  redshifts of 10,000 \kms.}
\end{figure}
\section{Infrared photometry}
To source infrared imaging of higher resolution than 2MASS,
the positions of the MeerKAT \HI-detections were uploaded to the Wide Field Astronomy Unit of the Royal Observatory Edinburgh, and \K band matches of 1 arcmin square fields were received as fits files.
The positional accuracy of HIZOA galaxies is $\sim$ 4~arcmin. Galaxies offset by $>$ 4$^\prime$ were excluded from this dataset. 
Both ESO's VISTA Hemisphere Survey and Via Lactea Survey were drawn on (Minniti \etal 2010, McMahon \etal 2013). 

The position angle and ellipticity were measured for each object in Table 1 that was not clearly face-on, using the algorithm employed by Mould \etal (2024) for WALLABY project data.
Total magnitudes were obtained using the a-phot program (Merlin \etal 2019). Many of the fields were so crowded with foreground stars that a star removal algorithm was employed\footnote{https://github.com/code2k13/starrem2k13/}. As this was an 8-bit astrophotography code, some scaling was required to preserve dynamic range\footnote{This was produced by sky subtraction together with dropping (and later restoring) the 3 least significant bits.}.
An upside of the dense foreground stellar population was that, even within these limited fields, ample 2MASS stars were present to serve as standards.

The resulting data for the 
\HI-detected galaxies and the here acquired \K ~band photometry are presented in Table 1, where values of the extinction A$_K$ are provided from IRSA's calculator\footnote{
https://irsa.ipac.caltech.edu/applications/DUST/}. Those from 
Schlafly \& Finkbeiner (2011) were chosen. 

\subsection{2MZoA}
For HIZOA galaxies (including the northern and Galactic Bulge extension) we used \K ~band counterparts in the 2MASX ZOA bright galaxy catalog compilation by Schr\"oder, van Driel \& Kraan-Korteweg  et al. 2019 ($K_s^o~ <$ 11.25 mag) that formed the basis of the 2MASX Redshift Survey (2MRS: Huchra \etal 2012, Macri \etal 2019), as well as its recent extension to fainter extinction-correction magnitudes ($K_s^o <$ 11.75 mag; Schr\"oder, 2025;
see Table 2). Assisted by complementary photometric and redshift data, when available, the likelihood of extended sources being galaxies were classified 
in those compilations (i.e. column 4, her Table A.1). 

The combined data set of our 21 cm ZOA galaxy selection contains 282 sources, omitting a few with discrepant redshifts or absent 'galaxy' flags. Velocities in the right hand column are in the cosmic microwave background reference frame. 
ESO IDs (Lauberts \etal 1981) for 2MZoA galaxies are given in notes to Table 2.

\subsection{Additional photometry}
Tables 4, 5 \& 6 present photometry by Alonso \etal (2025),
Williams \etal (2014), and 
by Said \etal (2016b). 
Overall there are 309 additional galaxies, some 200 of which have K band photometry from multiple sources. The $rms$ difference between these data is 0.2 mag. There are also a number of galaxies with 50\% velocity widths from multiple sources. The $rms$ difference in this case is 5\%, which propagates to less than 10\% in distance. 
Over the whole ZOA compendium, 39\% of the galaxies come from the HIZOA catalog, 18\% from the GA (Great Attractor) catalog, 11\% from the Local Void catalog and 31\% from the Vela sample.
\section{The Tully-Fisher relation}
The Tully-Fisher relation (TFR) for the galaxies in Tables 1--6 
is shown in Fig.~3.  Velocity widths corrected for galaxy inclination\footnote{The axial ratios $b/a$ determine the inclination angle of the galaxy to the plane of the sky, $i$, according to $$\frac{(b/a)^2-0.04}{1- 0.04} = \cos^2 i {~~~\rm and~~~~} \Delta V(0) = w_{50}/\sin i$$} are shown as $\Delta$V(0), based on  axial ratios measured in the \K ~band. Our assumed uncertainties for 2MZOA photometry, similar to those in the 2MRS survey, are lower limits in ZOA crowded fields.
Twenty-nine galaxies are in common with
the $Cosmic~Flows~4$ data base. Comparison of the TFR distances in CF4 with TFR distances from Fig.~3 shows agreement within $\pm$10\%.

Following the practice in the $CosmicFlows$ project (Courtois \etal 2011), FWHM 21 cm line widths included in Tables 1--6 
are also corrected for redshift in Fig.~3. Neither inclination nor redshift has a measured uncertainty, so we took the mean of the ratio of the widths at 20\% (w1) and 50\% (w2) of the maximum flux (1.34 $\pm$ 0.03) and assigned an uncertainty of (w1/1.34 - w2)/2. The maximum error bar shown in Fig.~3 is 11\% of w2.
 Distances in Fig.~3 were calculated in the cosmic microwave background 
(CMB) velocity frame assuming a Hubble Constant of 73 \kms~Mpc$^{-1}$ (Lineweaver \etal 1996, Riess \etal 2024).
Galaxies are colour-coded by this distance divided by 1000 \kms, and
the sequence is shown just above the x-axis of the plot.
\begin{figure*}[h]
\vspace{-2cm}
	\includegraphics[width=1.1025\textwidth]{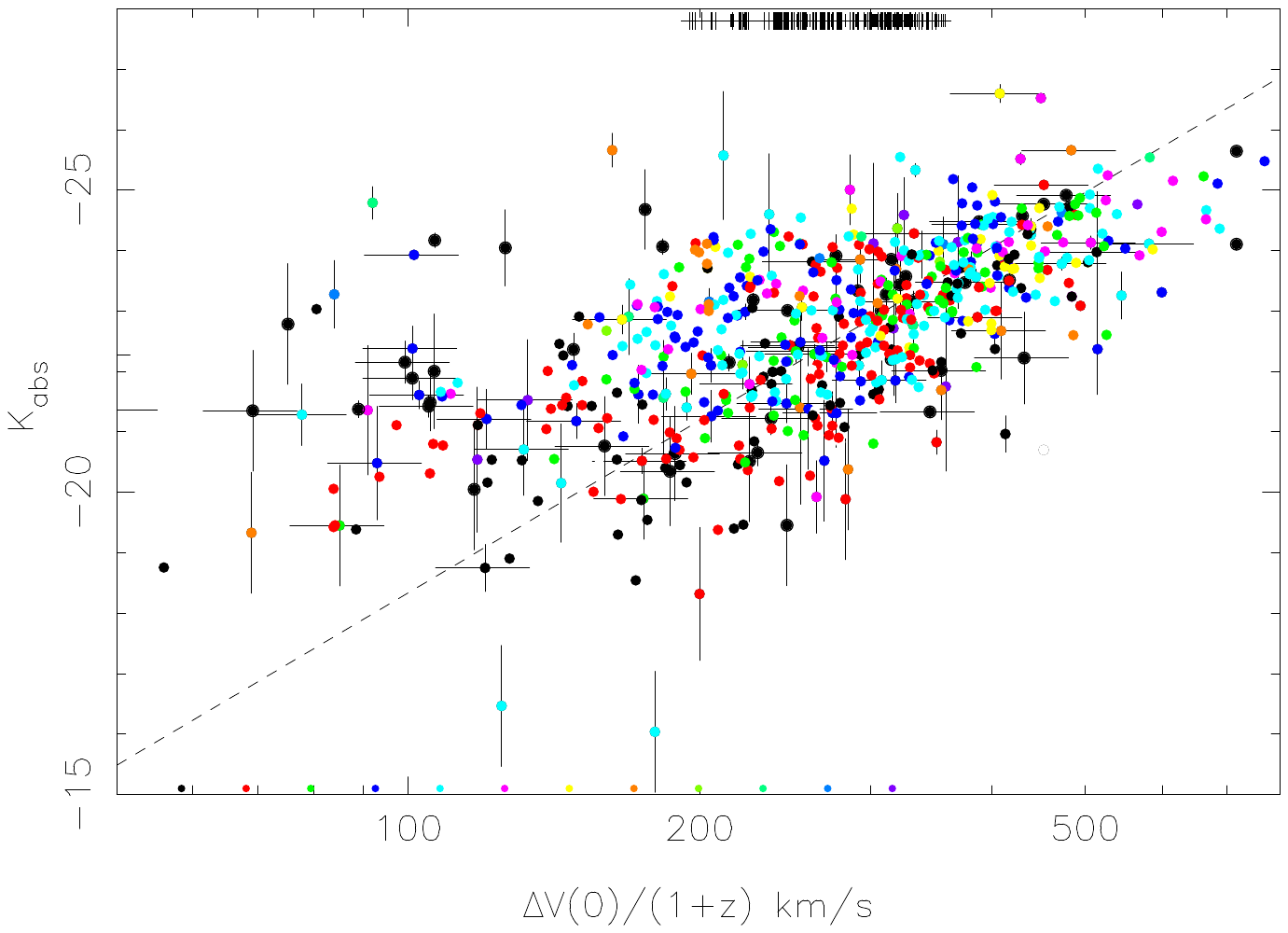}
	\caption{Tully-Fisher relation for the galaxies in Tables 1 and 2.
Galaxies in the ZOA are colour coded by their distance divided by 1000 \kms, and
	the sequence is shown just above the x-axis. The dashed line is
	from Masters, Springob \& Huchra (2014). 
The range in Galactic longitude
of these galaxies, running from 200 -- 360$^\circ$ is indicated by the concentration of markers on the upper horizontal axis.}
\end{figure*}

We also analyzed infrared photometry of a high latitude dataset by Ponomareva \etal (2021). 
The $rms$ scatter in the TFR was 0.95 mag
of which 0.70 mag was the photometric error
given by a-phot.
\section{Results and analysis}
After Malmquist bias correction of 1.38$\sigma^2$, where $\sigma$ accounts for observational uncertainties, 
Table~7 presents the mean departure in magnitudes of the grouped galaxies
from the dashed line, after exclusion of deviates more than 3 mag from the line.
\setcounter{table}{6}
\begin{table}[H]
	\caption{Distance modulus residuals}
    \begin{center}
\begin{tabular}{lrrr}
v&$\delta$K&$\pm$&n\\   
\kms&mag&mag&\\
\hline
  1000&  0.26 & 0.14&   58\\
  2000 &-0.00&  0.12 & 122\\
  3000&  0.16  &0.09  & 71\\
  4000 &-0.40 & 0.13  &104\\
  5000& -0.14&  0.14&  117\\
  6000 &-0.36  &0.22 &  40\\
  7000& -0.07  &0.23  & 30\\
  8000 &-0.38 & 0.40   &21\\
  9500 &-2.28&  1.18    &17\\
\hline

 \end{tabular}
 \end{center}
\end{table}
There is only one statistically significant distance modulus residual from 1000--8000 \kms.
At 4000 \kms~ the detected mean peculiar velocity is 800 $\pm$ 300 \kms. 
However, this is concentrated between Galactic longitudes 300--330$^\circ$.
The last row of Table 7, the most distant, may be most affected by Malmquist bias and is not considered significant.
\subsection{Selection effects}
The overall MeerKAT ZOA survey is HI limited and optically blind. So the sample is limited to gas rich galaxies, and progressively more so for more distant galaxies. The ALFALFA survey is similar in this regard, and Toribio \etal (2011) find very little correlation between HI content and position in the TFR. A change of an order of magnitude in the HI mass would make a 0.09 change in the red magnitude.

Galaxies that are lost to our sample due to uncertainty in the MeerKAT position are significant in number. The 1$^\prime$ ``postage stamp" in Fig. 1 shows that a positional error of more than 1$^\prime$ would cause the galaxy to be missed from the IR photometry in Table~1. Images of 2$^\prime$ could have been obtained, but the possibility of misidentification rises with increased error radius. Missing galaxies from this cause does not affect the TFR, but simply decreases the efficiency of our survey.

Galaxies can also be missed because of high extinction, but this is also not a source of bias, as higher foreground extinction makes for higher K$_0$ uncertainty.

Figures 4 \& 5 are redshift cuts and longitude cuts of Fig. 3. The separation of the high velocity and low velocity points in Fig. 5 shows the effect of Malmquist bias in favoring high luminosity deviant galaxies.
\section{Discussion}
There clearly are large structures located behind the Galactic plane.
The Laniakea supercluster (Tully \etal 2014) is a major structure,
and the Norma (Kraan-Korteweg et al. 1996, Woudt et al. 2004) and Vela (Kraan-Korteweg et al. 2017) 
 superclusters are associated.
The question we are asking here is, what is the dynamical significance
of these hidden masses? The largest estimate is for the Quipu structure
at 2.4 $\times$ 10$^{17}$ M$_\odot$ (B\"ohringer \etal 2025). At an average  distance
of 176 Mpc this would create an infall peculiar velocity of 150 \kms at a distance of 100 Mpc
away from it. In our direction that would be an influence at a redshift of 5500 \kms,
right where our sample is concentrated. 
\begin{figure}
    \hspace{-1cm}
    \includegraphics[width=1.3\linewidth]{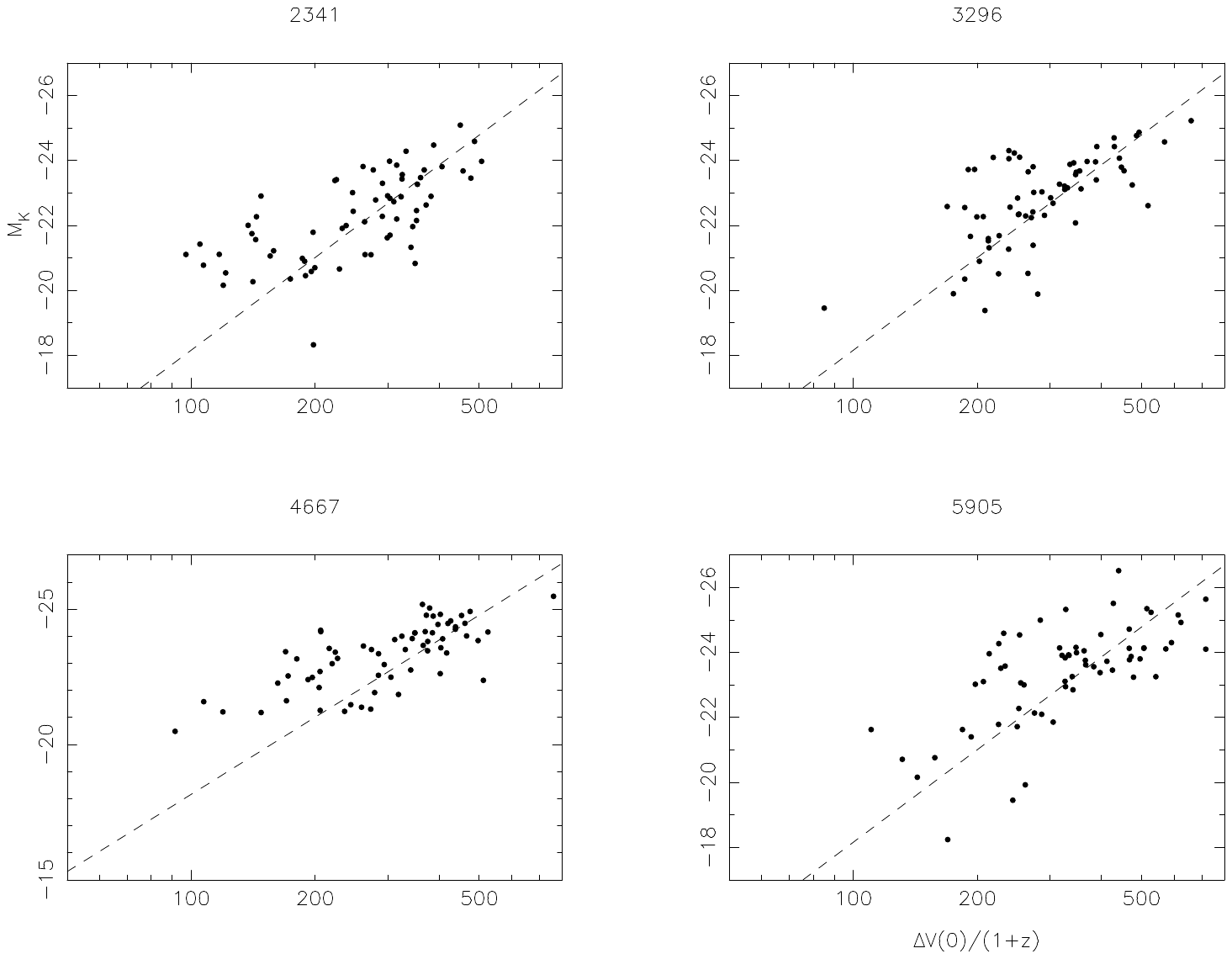}
    \caption{TFRs in redshift quartiles. The median redshift is above each plot.}
    \label{fig:placeholder}
\end{figure}
\begin{figure}
    \hspace{-1cm}
    \includegraphics[width=1.3\linewidth]{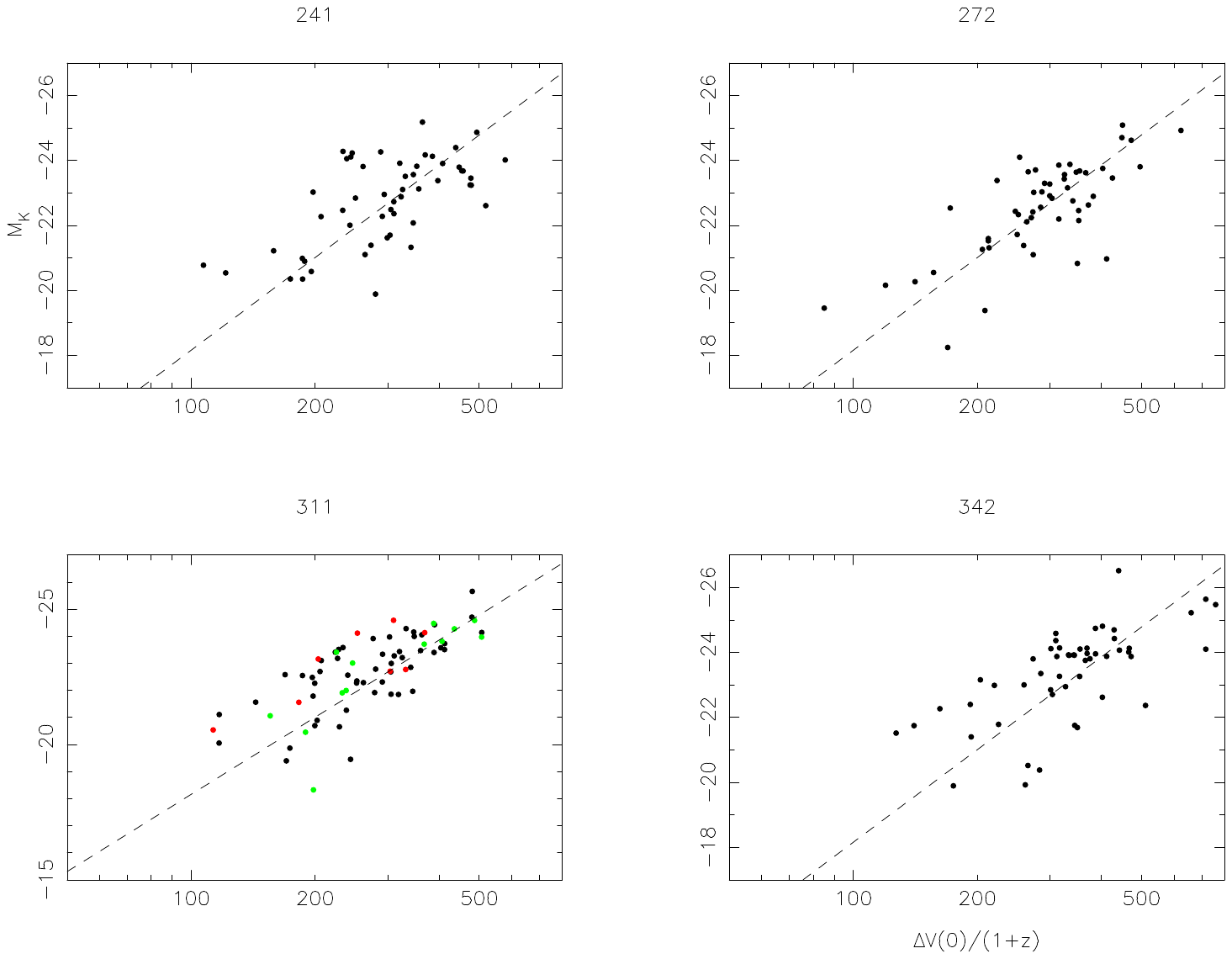}
    \caption{TFRs in Galactic longitude quartiles. The median longitude is above each plot. The 3rd quartile panel has red points for cz  $>$ 10$^4$ \kms and green for cz $<$ 3000 \kms.}
    \label{fig:laceholder}
\end{figure}
Our results suggest a velocity field in the Great Attractor region of the ZOA
consistent with such a  structure, with fractional peculiar
velocities of $\delta$v/cz $\lesssim$ 0.1. Hollinger \etal (2026) are
carrying out a reconstruction of this region, using the plentiful
redshifts from MeerKAT without redshift independent distances from the
Tully Fisher relation, of galaxies that are invisible even in the infrared.
The results from the infrared detected subset presented here are supportive of this approach.

\section{Conclusions}

We find a velocity field in the Great Attractor region of the ZOA
with fractional peculiar
velocities of $\delta$v/cz $\lesssim$ 0.1. 
This permits reconstruction of the region without recourse to redshift independent distances. The region remains of particular interest to cosmography, but also to cosmology, because of the continuing tension between the bulk flow predicted by $\Lambda$CDM and velocity field measurements out to z = 0.1.

Over 9000 galaxies have been included in the Hollinger \etal reconstruction of the ZOA octant between the Vela and Shapely superclusters using redshifts alone. Even larger cosmology projects are possible with the strategy we have described here. DESI DR1 (Carr, Howlett \& Ansellem 2025) have measured 100,000 peculiar velocities from the 18.7 million galaxy redshifts that have been obtained. 

A mass map of an adequate and complete volume is essential, if the tension between the large cosmic microwave dipole and the expectations of the standard cosmology are to be understood.

\section*{References}


\noindent 
\label{sec:xxx}
Aaronson, M., Huchra, J. \& Mould, J. 1979, ApJ, 229, 1\\
Alonso, M. \etal 2025, A\&A, 700, 33\\
B\"ohringer, H. \etal 
 2025, A\&A, 695, A59\\
 Carr, A., Howlett, C. \& Ansellem, A. 2025, arxiv 2502.03232\\
Chamaraux, P.  \etal 1990, A\&A, 229, 340\\  
Courtois, H.~M. \etal 2011, MNRAS, 414, 2005\\
Courtois, H.~M. \etal 2013, AJ, 146, 69\\
Courtois, H.~M. \etal 2025, A\&A, 701A, 187\\
Donley, J. \etal 2006 MNRAS, 369, 1741\\
Dressler, A., Lynden-Bell, D. \& Burstein, D. 1987, ApJ, 313,~42\\
Dupuy, A., Courtois,  H.~M., 2023, A\&A, 678, 176\\
Feldman, H., Kaiser, N. \& Peacock, J. 1994, ApJ, 426, 23\\
Focardi, P. \etal 1984, ASSL, 111, 55\\
Goedhart, S. \etal 2024, MNRAS, 531, 649\\
Hasegawa,  T. \etal 2008, MNRAS, 316, 326\\
Hatamkhani, N., Kraan-Korteweg, R., Blyth, S.,\& Skelton, R. 2024 ApJ, 972, 57\\
Hollinger, A., Courtois, H.~M., Mould, J., Rajohnson, S. \& Kraan-Korteweg, R. 2026, arxiv.org/2603.09339\\
Huchra, J. \etal 2012, ApJS, 199, 26\\
Jarrett, T. H. \etal 2000, AJ, 120, 298\\
Kolatt, T., Dekel, A. \& Lahav, O. 1995 MNRAS, 275, 797\\
Koribalski, B. \etal 2020, MNRAS, 478, 1611\\
Kraan-Korteweg, R.C. 2005, Rev. in Modern Astr, 18, 48\\
Kraan-Korteweg et al. 1996, Nature, 379, 519\\
Kraan-Korteweg, R.C. \etal 2008, ASSP 5, 13\\
Kraan-Korteweg, R.C. \etal 2017, MNRAS 466, L20\\
Kraan-Korteweg, R.C. \etal 2024, IAUGA, 32, 893\\
Kraan-Korteweg, R.C. \etal 2026, IAU S392 (in press)\\
Kurapati, S. \etal 2024, MNRAS, 528, 542\\
Lauberts, A. \etal 1981, A\&AS, 43, 307\\
Louw, A. \etal 2026, IAU S392 (in press)\\
Kurapati, S. \etal 2024, MNRAS, 528, 542\\
McMahon, R., Banerji, M., Gonzalez, E. \etal 2013, The Messenger, 154, 35\\
Macri, L. \etal 2019, ApJS, 245, 6\\
Masters, K., Springob, C. \& Huchra, J. 2014, AJ, 147, 124\\ 
Merlin, E. \etal 2019, A\&A, 622, 169\\ 
Minniti, D. \etal 2010, New Astron., 15, 433\\
Mould, J. \etal 2024, MNRAS, 533, 925\\
Mould, J. 2026, A\&A, in press, arxiv 2601.07106\\
Nusser, A. 2025, arxiv 2512.04075\\
Ponomareva, A. \etal 2021, MNRAS, 508, 1195\\
Rajohnson, S. \etal 2024b, MNRAS, 535, 3429\\
Rajohnson, S. \etal 2024a, MNRAS, 531, 3486\\
Ramatsoku, M. \etal 2016, MNRAS, 460, 923\\
Ramatsoku, M. \etal 2020, A\&A, 644, A107\\
Riess, A. \etal 2024, ApJ, 977, 120\\
Said, K. \etal 2016a, MNRAS, 457, 2366\\
Said, K. \etal 2016b, MNRAS, 462, 3386\\
Schlafly, E. \& Finkbeiner, D. 2011, ApJ 737, 103\\
Schr\"oder, A., van Driel, W. \& Kraan-Korteweg, R.  2019, MNRAS, 482, 5167\\
Schr\"oder, A. 2025, A\&A, 704, 252\\
Shaya, E., Tully, R. B., Pomar\`ede, D. \& Peel, A. 2022 ApJ, 927, 168\\ 
Staveley-Smith, L. \etal 2016, AJ, 151, 52\\
Steyn, N., 2003, MSc thesis UCT \\
Steyn, N. \etal 2024 MNRAS, 529, L88\\
Steyn, N. \etal 2026, IAU S392 (in press)\\
Toribio, M. C. \etal 2011, ApJ, 732, 93\\
Tully, R. B. \etal 2008, ApJ, 676, 184 \\
Tully, R. B., Courtois, H., Hoffman, Y. \& Pomar\`ede, D. 2014, Nature, 513, 71 \\
Tully, R. B. \& Fisher, J.R. 1977, A\&A, 54, 661\\
Wakamatsu, K. et al. 1997, PASA, 14, 126 \\
Williams, W., Kraan-Korteweg, R. \& Woudt, P. 2014 MNRAS, 443 41\\
Woudt, P.A. et al. 2004, A\&A, 415, 9\\

\section*{Acknowledgements}
We thank Nadia Steyn, Austun Louw and Anja Schr\"oder for providing data in advance of its availability at CDS. The MeerKAT telescope is operated by the South African Radio Astronomy Observatory, which is a facility of the National Research Foundation, an agency of the Department of Science and Innovation. This publication made use of data products from the Two Micron All Sky Survey, which is a joint project of the University of Massachusetts and the Infrared Processing and Analysis Center, funded by the U.S. National Aeronautics and Space Administration (NASA) and the National Science Foundation.
The research also made use of the NASA/IPAC Infrared Science Archive, which is funded by NASA and operated by the California Institute of Technology.  HMC and JRM acknowledge support from the Institut Universitaire de France and the Australian Research Council respectively.
\section*{Data Availability}
https://github.com/jrmould/zoa/
\setcounter{table}{2}
\begin{table}[H]
\caption{ESO galaxy IDs in Table 2}
\begin{tabular}{|llll|}
\hline
$^0$ESO429-9&
$^1$ESO429-8  &      
$^{24}$ESO429-14&
$^2$ESO493-6      \\  
$^{16}$ESO493-24&
$^3$ESO494-12     &  
$^4$ESO494-25       &
$^5$ESO430-20       \\
$^6$ESO494-29       &
$^8$ESO431-1        &
$^{25}$ESO370-15&
$^9$ESO371-3      \\  
$^7$ESO314-3 &
$^{10}$ESO261-6&        
$^{22}$ESO213-6&
$^{11}$ESO213-9  \\      
$^{23}$ESO92-22 &
$^{12}$ESO173-15  &     
$^{13}$ESO97-12     &   
$^{18}$ESO176-6\\
$^{19}$ESO274-21&
$^{14}$ESO330-11  &     
$^{15}$ESO390-4     &   
$^{21}$ESO395-4\\
$^{20}$ESO395-6  &&&\\
\hline
\end{tabular}
\end{table}      

\vfill\break
\clearpage
\pagebreak
\setcounter{LTchunksize}{256}
\setcounter{table}{0}
\begin{longtable}{lrrrrrrrr}
\caption{ZOA galaxies}\\ 
\hline
Name&$l$&$b$&w50&K&$\pm$&A$_K$&v$_{CMB}$&b/a\\
&deg&deg&km&mag&mag&mag&km&\\
&&&s$^{-1}$&&&&s$^{-1}$&\\
\hline
\endfirsthead
\multicolumn{9}{l}
{\tablename\ \thetable\ -- \textit{Continued}}
\endhead
\hline \multicolumn{9}{r}{\textit{Continued on next page}}\\
\endfoot
\hline
\endlastfoot
HIZOA J0653-03B   &  216.62&   -1.35&  283&   13.53&    1.00&  0.42&   2900&  0.20 \\
065919-013114     &  215.18&    1.06&  171&   12.15&    0.21&  0.24&   2069&  0.30 \\
070019-022344     &  216.08&    0.88&   50&   11.37&    0.38&  0.26&   2115&  0.20 \\
070934-052541     &  219.81&    1.55&  273&   11.10&    0.09&  0.14&   2092&  0.64 \\
071248-091823     &  223.65&    0.46&   84&   11.78&    0.14&  0.21&   2832&  0.40 \\
071354-071708     &  221.97&    1.63&  189&   13.51&    0.90&  0.16&   4007&  0.20 \\
072211-055546     &  221.71&    4.09&  263&    8.45&    0.03&  0.08&   1997&  0.20 \\
072513-265859     &  240.67&   -5.15&  422&   10.88&    0.13&  0.13&   7805&  0.40 \\
\end{longtable}
\vfill\break
\pagebreak
\clearpage
\setcounter{table}{1}
\begin{longtable}{lrrrrrrrr}
\caption{\bf Galaxies from the 2MZoA catalog to be truncated in published version}\\
\hline
ID      &    RA  &   Dec  &   $l$    &     $b$  &    K$_0$& w50&  b/a & v$_{CMB}$\\
        &   deg  &    deg &  deg   &   deg  &    mag & \kms&      & \kms\\
\hline
\endfirsthead
\multicolumn{9}{l}
{\tablename\ \thetable\ -- \textit{Continued}}
\endhead
\hline \multicolumn{9}{r}{\textit{Continued on next page}}\\
\endfoot
\hline
\endlastfoot
J0622+04&  95.621&   4.532&  205.58&   -4.34&   10.02&  306&  0.40&  2777\\
J0623+14&  95.829&  14.728&  196.66&    0.61&   10.08&  286&  0.68&  5167\\
J0623+04&  95.900&   4.382&  205.84&   -4.17&    9.85&   97&  0.20&  2688\\
J0630+16&  97.521&  16.844&  195.54&    3.03&   11.48&  258&  0.32&  2397\\
J0635+11&  98.975&  11.162&  201.24&    1.67&   10.84&  408&  0.66&  3572\\
J0635+14&  98.988&  14.598&  198.19&    3.25&   10.06&  329&  0.40&  3872\\
J0643+06& 100.825&   6.966&  205.81&    1.37&   11.44&  356&  0.52&  6720\\
J0644+12& 101.004&  12.433&  201.01&    4.02&    9.88&  403&  0.52&  3765\\
\end{longtable}
\setcounter{table}{3}
\vfill\break
\pagebreak
\clearpage
\begin{longtable}{lrrrrrrrr}
\caption{Williams \etal (2014) galaxies}\\
\hline
Name&$l$&$b$&w50&K&Note&A$_K$&v$_{CMB}$&b/a\\
	&deg&deg&km&mag&&mag&km&\\
    &&&s$^{-1}$&&&&s$^{-1}$&\\
    \hline
    \endfirsthead
\multicolumn{9}{l}
{\tablename\ \thetable\ -- \textit{Continued}}
\endhead
\hline \multicolumn{9}{r}{\textit{Continued on next page}}\\
\endfoot
\hline
\endlastfoot
  J0622+11B       &  199.78&   -1.15&  366&   11.02&&  0.32&   5594&  0.38 \\
  J0709-05        &  219.83&    1.54&  248&    9.49&&  0.14&   1903&  0.64 \\
  J0716-18C       &  232.57&   -3.11&  253&   10.77&&  0.54&   3002&  0.20 \\
  J0727-23        &  238.23&   -3.31&  398&   10.42&&  0.33&   4600&  0.34 \\
  J0730-22        &  236.82&   -1.85&  262&    9.20&&  0.57&    987&  0.20 \\
  J0740-22        &  238.83&   -0.10&  155&   11.17&&  0.18&   3276&  0.70 \\
  J0740-32        &  246.80&   -4.61&  264&   10.20&1&  0.20&   2927&  0.36 \\
  J0741-22        &  238.56&    0.24&  452&    9.68&&  0.20&   3296&  0.20 \\
\end{longtable}
\begin{table}
\begin{tabular}{|llllll|}
\hline
Note&Galaxy&Note&Galaxy&Note&Galaxy\\
           1 &ESO429-12 &  
           2 &ESO560-12  & 
           3 &ESO560-13   \\
           4 &ESO493-16 &  
           5 &ESO493-24  & 
           6 & ESO494-26  \\
           7 &ESO371-11 &  
           8 &ESO314-4   & 
           9 &ESO213-3    \\
          10 &ESO127-11 &  
          11 &ESO213-9   & 
          12 &ESO168-9    \\
          13 &N3882    &   
          14 &ESO173-IG5&  
          15 &ESO173-15  \\ 
          16 &ESO97-12    &
          17 &ESO175-4    &
          18 &ESO175-9    \\
          19 &ESO170-6    &
          20 &ESO329-19   &
          21 &ESO330-11   \\
          22 &ESO330-10   &
          23 &ESO137-14   &
          24 &ESO390-4    \\
          25 &ESO452-3    &
          26 &ESO517-7    &
          27 &ESO452-7    \\
          28 &ESO179-12   &
          29 &ESO278-7    &
          30 &ESO395-2    \\
          31 &ESO395-4    &
          32 &ESO395-6    &
          33 &ESO457-15   \\
\hline
\end{tabular}

\vspace{1cm}
\end{table}

\begin{table*}
\caption{Alonso \etal (2025) galaxies}
\begin{tabular}{lrrrrrrr}
\hline
Name&$l$&$b$&w50&K$_0$&v$_{CMB}$&b/a\\
&deg&deg&km&mag&km&\\
&&&s$^{-1}$&&s$^{-1}$&\\
\hline
  J0635+14B       &  198.18&    3.25&  329&   10.25&   4125&  0.28 \\
  J0738-16        &  233.31&    2.21&  272&   11.26&   3392&  0.66 \\
  J0746-18        &  235.66&    3.18&  361&   10.18&   3539&  0.69 \\
  J0748-25A       &  241.65&    0.28&  199&   11.13&   8156&  0.64 \\
  J0748-26        &  242.59&   -0.24&  226&    8.92&   2873&  0.40 \\
\end{tabular}
\end{table*}
\vfill\break
\pagebreak
\clearpage
\begin{longtable}{lrrrrrrr}
\caption{Said \etal (2016b) galaxies}\\
\hline
Name&$l$&$b$&w50&K$_0$&v$_{CMB}$&b/a\\
&deg&deg&km&mag&km&\\
&&&s$^{-1}$&&s$^{-1}$&\\
\hline 
\endfirsthead
\multicolumn{8}{l}
{\tablename\ \thetable\ -- \textit{Continued}}
\endhead
\hline 
\multicolumn{8}{r}{\textit{Continued on next page}}\\
\endfoot
\hline
\endlastfoot
  J0635+14A           &  197.78&    3.30&  275&   10.62&   3941&  0.27 \\
  J0652-03        &  216.27&   -1.55&  173&   12.06&   2772&  0.35 \\
  J0653+07        &  206.80&    3.71&  329&   12.65&   7185&  0.29 \\
  J0657-13        &  225.40&   -4.61&   69&   13.00&   6396&  0.68 \\
  J0658-12        &  224.75&   -4.06&  113&   12.83&   5670&  0.26 \\
  J0659-00        &  214.27&    1.69&  264&   12.88&   7152&  0.64 \\
  J0700-11        &  224.51&   -3.27&  410&    9.72&   2918&  0.30 \\
  J0702-03A       &  217.04&    0.93&  402&   11.49&   6857&  0.53 \\
\end{longtable}
\begin{table}
\begin{tabular}{l}
    * 
\end{tabular}
\end{table}
\end {document}